\documentstyle[numreferences]{kapproc}

\begin{opening}
\title{Classification of Multivector Theories and
the Modification of the Postulates of Physics}
\author{William M. Pezzaglia Jr.}
\institute{Department of Mathematical Sciences\\
St. Marys College of California\\
PO Box 3517\\
Moraga, CA 94575\\U.S.A.}
\date{(Received: May 1993)}
\end{opening}
\runningtitle{Preprint\# clf-alg/pezz9303}
\begin{document}

\begin{abstract}

We propose a graded classification of the entire field of multivector
physics, including all alternative points of view.  The (often tacit)
postulates of different types of formulations are contrasted, summarizing
their consequences.  Specifically, spin-gauge formulations of gravitation
and GUT which assume standard column spinors will require unnecessarily
large matrix algebras.  An extreme generalization is introduced, where
wavefunctions are assumed to be multivectors, in which multiple
generations of particle families naturally appear without increasing
the size of the algebra.  Further, this allows for two-sided (bilateral)
operators, which can accomodate in excess of 10 times more gauge fields
without increasing the algebraic representation.  As this generalization
encompasses all the essential features of the other categories, it is
proposed to be the best path to new physics.

Summary of Talk given May 4, 1993 at Deinze, Belgium,
To appear in: Procceedings of the Third
International Conference: Clifford Algebras and Their Applications
in Mathematical Physics, R. Delanghe, F. Brackx and H. Serras Ed.,
University of Ghent, Belgium (Kluwer 1993).  ISBN 0-7923-2347-5
\keywords{} spin-gauge,  multivector,  clifford
\end{abstract}

\section{Introduction}
Collectively, all work in this field can be divided into two broad categories:
{\it Mathematics}  and {\it Physics}
(and the relationships between these disciplines).  Before we concentrate
on the latter, we note that the former can be further
subdivided into two areas, {\it Algebra}  and {\it Analysis}.
The structure of Clifford algebra, its relation to other systems
(e.g. Grassmann, Lie or Cayley algebras) is the concern of an algebraist
with an ultimate goal of creating a single unified math language based on
geometry.  The analyst is concerned with the calculus of functions
of multivectors, with the potential of providing coordinate-free
formulations of fields in curved spacetime.

In contrast, only a small minority of published papers actually address
the role of using Clifford Algebras in physical models.  The category of
{\it multivector physics} can be subdivided into two areas,
{\it Philosophy} and {\it Formulation}.
Generically, the former is concerned with guiding principles, which has
been ignored by all but a few authors.
This is unfortunate because basic postulates can not be deduced from logical
reasoning or pure math (e.g. from the Clifford algebra).  There has been more
attention to formulation, i.e. the association of physical phenomena with
mathematical symbols, however no "new" physics will be obtained by applying
Clifford algebra while tacitly assuming the old postulates.

The intent of this paper is to focus more upon the different underlying
philosophies involved in how multivector physics are formulated.  We
introduce two broad classifications for comparison and contrast,
organized along classic greek lines, the
{\it Platonistic} and {\it Aristotelian}.

\section{Platonistic}
Generally, standard theory has the classical Platonistic view that abstract
mathematical entities exist.  We are forbidden (due to the imperfection of
physical existence) to directly experience (i.e. "measure") these ideal forms
(e.g. "quantum phase").  They are only knowable through mathematical
formulation.  The physical structure of space is hence really
mathematical structure in disguise.

\subsection{Standard Theory}
Specifically, in quantum mechanics the fundamental structure of "real"
particles is represented by abstract multicomponent spinors which exist in
transcendental complex {\it spin space}, which has no concrete physical
analog.  The unobservable nature of spin space is built into the
postulates of quantum mechanics, wherein the
wavefunction's "phase" angle (in complex phasor space) is ordained to be
unmeasurable.   Hence the theory (e.g. the Lagrangian) must be formally
invariant with respect to changes in phase, which are generated by the
$U(1)$ Lie group.  This symmetry
{\it gauge transformation}  induces (via Noether's theorem) bilinear covariant
forms independent of the unobservable phase.  Obeying a conservation law,
these forms are associated via the correspondence principle with concrete
measurable quantities (e.g. $\psi^{\dagger}\psi $ is the
observable probability density).

Requiring the formulation to be invariant under local changes in phase
(different at each point in space) is called the
{\it principle of local gauge invariance}.  Through
an inductive generalization called {\it minimal coupling}, a gauge covariant
derivative is induced which contains a connection function interpreted to be
the
electromagnetic 4-vector potential.  The curvature of this connection is the
electromagnetic field strength tensor, whose source is found to be the bilinear
covariant current induced by the gauge transformation on the Lagrangian.

Requiring invariance with respect to larger non-abelian symmetry groups
induces other gauge fields e.g. "gluons" from $ SU(3)$.  The emphasis
of modern physics has shifted to choosing the right "magical" symmetry
group of the universe which will correctly describe the known four
fundamental forces in one "unified theory".
Generally the group is expressed as a complex matrix which operates on the
column spinor wavefunction.  The size of the representation induces the number
of degrees of freedom in the "spinor-space", which in order to incorporate all
the known fermions must unfortunately be in excess of 90.

\subsection{Operationalist}
The most conservative use of Clifford Algebra in quantum physics is to view its
elements only as abstract operators on a spinor space.  Only those
symmetry groups which are embedded within a global Clifford algebra are to
be considered for gauge field generators.
Unfortunately this restricts one to groups such as $SU(2^{d})$,
where $d$ is an integer, eliminating the standard $SU(3)$ and $SU(5)$ [except
as subgroups of a larger algebra].  Further, since the representation size
is restricted to $2d$, this forces global multispinors of either probably
two few (64) or too many
(128) components for the estimated 90 or 96 known degrees of freedom needed to
represent all the fermions of the standard model.

One argument for using Clifford algebra is that it naturally appears in
relativistic quantum mechanics.  Four mutually anticommuting algebraic
entities were required to factor the Klein-Gordon
equation to the first order Dirac equation.  These four
elements generate the 16 element Majorana algebra, which has the lowest order
representation of 4 by 4 matrices, inducing 4 component {\it bispinor}
wavefunctions.  These bispinors successfully correspond to a Dirac fermion
(e.g. electron) with "half-integral spin".  In order for the formulation to
be generally covariant, the 4 basis generators obey the defining condition
of a Clifford algebra,
${1 \over 2} \{ \gamma_{\mu},\gamma_{\nu} \}=
g_{\mu \nu}= {\bf e}_{\mu} \bullet {\bf e}_{\nu}$
where ${\bf e}_{\mu}$ is a coordinate basis vector (not part of a Clifford
algebra).
Hence the $\gamma_{\mu}$ can be determined from the Riemann space metric
$g_{\mu \nu}$ up to a similarity transformation,
$\gamma_{\mu}'= S\gamma_{\mu} S^{-1}$, i.e. a change in matrix representation,
called a {\it spin transformation}.

{\it Spin gauge theories} \cite{Chis89} invoke the principle of
{\it local representation invariance}, i.e.
require the theory to be invariant under similarity
transformations which are a function of position.  This induces a spin
covariant derivative $\nabla_{\mu}$ where the {\it spin connection}:
${\Omega_{\mu} = \Omega_{\mu}^{\ i} \Gamma_i}$ (Fock-Ivanenko coefficient) is
interpreted to be a Clifford aggregate ($\Gamma^i$ is an element of the
algebra) of new spin gauge fields,

$$ \psi' = S \psi ,\eqno(1a)$$
$$ \nabla_{\mu} \psi^{\alpha} = \partial_{\mu} \psi^{\alpha} +
\Omega_{\mu}^{\ \alpha \beta} \psi_\beta  ,\eqno(1b)$$
$$ \Omega_{\mu}' = S\Omega_{\mu}S^{-1} + (\partial_{\mu}S)S^{-1} ,\eqno(1c)$$
$$ K_{\mu \nu} = [\nabla_{\mu},\nabla_{\nu}] =
\Omega_{\mu;\nu} - \Omega_{\nu;\mu} + [\Omega_{\mu},\Omega_{\nu}] =
K_{\mu \nu}^{\ i} \Gamma_{i} ,\eqno(1d)$$
$$ \gamma^{\nu}_{;\mu} - [\Omega_{\mu},\gamma^{\nu}]=0 \eqno(1e)$$

The last equation, requiring the spin covariant derivative of any element of
the Clifford algebra to vanish is arbitrarily imposed.  It is a sufficient
(but not necessary) condition to insure $g_{\mu \nu ;\alpha}=0$,
i.e. the coordinate covariant derivative of the metric
to vanish defining a Riemann space.  If it is desired that
$ \gamma^{\nu}_{;\mu}$ is grade preserving, then the spin connection, and the
resulting field strength tensor $K_{\mu \nu}$ must be a bivector (or the
commuting center of the algebra).

\subsection{Structuralist}
The connection between abstract spin space and tangible coordinate space can
be made more clear if we propose a geometric structure to spin a space.
A {\it spin transformation} can now be associated with a change of spin
basis.  A spin vector $\Psi= \psi^{\alpha} {\boldmath \xi_\alpha}$
will be manifestly invariant, where the basis spin vectors
${\boldmath \xi_\alpha}$ and spinor components $\psi^\alpha$
will be modified by the transformation.  The Fock-Ivanenko coefficients
$\Omega_{\mu}^{\ \alpha \beta}$, are now interpretable as the
spin space analogy of the Christoffel symbols, describing how the
spin vectors change with position in coordinate space,
$\partial_\mu {\boldmath \xi_\alpha}  =
{\boldmath \xi_\beta} \Omega_{\mu \enskip \alpha}^{\enskip \beta}$.
A non-vanishing field strength tensor $K_{\mu \nu}$ (eq. 1d) is now
interpretable as intrinsic spin space curvature.  If one demands that the
spin metric
$\eta_{\alpha \beta}= \eta^*_{\alpha \beta}=
{\boldmath \xi_\alpha}^{\dagger} {\boldmath \xi_\beta}$
be coordinate independent then it follows that the spin connection must be
restricted $\Omega_{\mu}^{\dagger}= - \Omega_{\mu}$.  In the particular case of
Minkowski spacetime with the associated Majorana Clifford algebra the spin
connection would be restricted to Dirac bar-negative elements, the vectors and
bivectors, which are generators of the Poincare group.  Hence it is tempting to
pursue this as a gauge derivation of gravity \cite{Craw93}.

An element $\gamma^{\mu}_{\alpha \beta}$ of the matrix Clifford algebra
is viewed as a {\it second rank spinor}.  The manifestly spin representation
independent bilinear form ${\bf e}^{\mu}=
{\boldmath \xi_\alpha} \gamma^{\mu}_{\alpha \beta}
{\boldmath \xi_\beta}^{\dagger}$
can be concretely interpreted as the observable basis vector of Riemann
space.  This forces a relationship between spin and coordinate space, in
particular eq. (1e) is built into this definition.  More importantly, this
will force a decomposition of the spin gauge field when the Clifford
gradient replaces the standard in Lagrangians and equations of motion.
For example, consider the non-relativistic Pauli-Schrodinger equation
for a two-component spinor.  The associated Clifford geometry of three
dimensional space is the Pauli algebra.  The generators of the unitary spin
transformations form the $U(1) \otimes SU(2)$ group with elements
$\Gamma^j=\{ i,i\sigma_1,i\sigma_2,i\sigma_3 \}$.  The induced spin connection
$\Omega_\mu = \Omega_\mu^k \Gamma_k =i A_\mu + i\sigma_k Z_\mu^k$
apparently contains the usual electromagnetic vector field $A_\mu$ plus
three other vectors \cite{Enja93}.  However, it is
$\sigma^k \nabla_k$ that appears in the Lagrangian (and hence equation
of motion) which forces a decomposition of the gauge fields,

$$-i \sigma^k \nabla_k = -i {\boldmath \nabla} +
{\boldmath A} + \phi + i{\bf C} ,\eqno(2a)$$
$$ {\boldmath A} = A^k \sigma_k ,\eqno(2b)$$
$$ \phi = Z_k^k  ,\eqno(2c)$$
$$ C^n = \epsilon^{jkn} Z_{jk} .\eqno(2d)$$

Apparently the symmetric part of the spin field $Z^{jk}$ does not directly
couple to the particle!  The remaining components enter as scalar and
pseudovector interactions (e.g. analogous to $f_0$ and $f_1$ mesons
respectively).  The sources of these fields will be spin currents;
in particular it is interesting to note that the
scalar field $\phi$ couples to the helicity density:
$J_\phi =i\psi^\dagger \sigma^k \partial_k \psi$ .

\section{Aristotelian}
In contrast, consider an Aristotelian view that abstract entities
(e.g. spin space) do not exist, rather the structure of particle fields
is (represented by) tangible physical geometry.  Both operators and
structure will be described by a single unifying, geometrically
interpretable Clifford Algebra.  We make a distinction as to the degree
by which the theory is committed to the full exploitation of the
geometric algebra.

\subsection{Pragmatism}
The {\it pragmatist}  notes that restricted combinations of the algebra
called minimal ideals can be used to replace the column spinor.  For example,
in the Pauli equation the basis spin vectors of a two component spin space
could be replaced by
$ {\boldmath \xi_2} = \sigma_1{\boldmath \xi_1}$ and
${\boldmath \xi_1}  = {1 \over 2}(1 + \sigma_3)$.
The quadratic form of the wavefunction
$\psi^{\dagger}\psi$ may no longer be a real scalar, but have other
geometric pieces, however the pramatist does not feel obligated to
utilize these extra pieces.  In fact, he may restrict the
geometric content of the wavefuntion such that the non scalar portions
disappear.  For example, consider a Dirac-Hestenes equation of the form
$\gamma^{\mu} \partial_{\mu} \psi = m\psi \Gamma $, where
$\Gamma$, is an element of the algebra such that
$\Gamma^2=+1$ if in the $(-+++)$ metric, else
$\Gamma^2=-1$ for the $(+---)$ metric).  For $\tilde{\Gamma} = -\Gamma$,
(the tilda represents the reversion anti-involution)
the multivector current equation is,
$$ \partial_{\mu} = m[ \tilde{\psi}\psi, \Gamma]  ,\eqno(3a)$$
$$j^\mu =\tilde{\psi} \gamma^\mu \psi  = s^\mu + T^{\mu \nu} \gamma_\nu +
\rho^\mu \gamma_0 \gamma_1 \gamma_2 \gamma_3 .	\eqno(3b)$$
For a unrestricted multivector wavefunction $\psi$ only the scalar current
$s^\mu$ is conserved, so the pragmatist could interpret it as an
observable and ignore the parts of eq. (3b) by taking the scalar part
of eq. (3a).  Alternatively, he would restrict the form of the wavefunction
such that the commutator of eq. (3a) vanishes, hence all three of the
currents of eq. (3b) are conserved and hence interpretable as observables.

Such formulations tend to be calculationally isomorphic to standard theory,
hence one can argue that it is mere reformulation.  One can adopt a
"minimalistic" principle, where the formulation uses the least number
of geometric degrees of freedom to represent the phenomena.  Hence
the postulates of quantum mechanics are mildly challenged in that
for example there is no imaginary "i" present in the 4 dimensional real
Clifford algebra.  Specifically the standard bias that the "i" is required
for charged fields is clearly not true.

\subsection{Radicalism}
In contrast, the {\it Radicalist} holds to the extreme position that in
formulation we are ontologically committed to use and interpret
EVERY geometric degree of freedom within the given dimension of the
space.  For example, consider the multivector Dirac
equation $\gamma^\mu \partial_\mu \psi = m \psi$ (only in the $-+++$ metric)
which has Greider \cite{Grei84} multivector current,
$$ \partial_\mu j^\mu = 0  ,\eqno(4a)$$
$$ j^\mu = \bar{\psi} \gamma^\mu \psi = T^{\mu \nu} \gamma_\nu +
M^{\mu \alpha \beta} \gamma_\alpha \wedge \gamma_\beta , \eqno(4b)$$
where the Dirac "bar" operator is the main algebra anti-involution
(reversion followed by inversion of the basis vector elements).

In contrast to eq. (3a), each part of the Greider current is conserved
(and hence interpretable as an observable) without any restriction
on the wavefunction $\psi$.  Hence the general multivector wavefunction
is no longer constrained to be a minimal ideal describing one particle,
but contains multiple generations of particles \cite{Pezz592}.  This
general structure may violate the standard Fierz identities \cite{Pezz692},
and introduce new quantum (hidden?) parameters.  A Lagrangian constructed
of the general multivector wavefunction may have non-scalar parts, which
could contribute in the path integral formulation of quantum mechanics.

Now allowable {\it dextral} (right-side applied)\cite{Pezz692} operators on
the wavefunction
must be included, which brings a challenge to the postulate of compatible
observables.  Two operators may not commute but will represent compatible
observables if one is a dextral operator while the other a {\it sinistral}
one (left-side applied, e.g. a spin-transformation).  Further, the same
operator when applied to the right has an entirely different physical effect
(observable) than when applied on the left.  For example, in the Pauli
algebra, the spin would be given by
$S_j=SP(\psi^{\dagger} \sigma_j \psi) = SP(\psi \psi^{\dagger} \sigma_j )$
while the isospin by
$T_j=SP(\psi^{\dagger} \psi \sigma_j ) = SP(\psi \sigma_j \psi^{\dagger} )$.
The SP notation means to take the scalar part (matrix trace) as the
fundamental bilinear covariant form $\psi^{\dagger} \sigma_j \psi $ will
in general no longer be a pure scalar, but a multivector.  Each
piece of such a multigeometric current must be given interpretation;
we cannot just pick the ones we like and exclude the rest.  In particular,
we have to consider new {\it bilateral currents} \cite{Pezz692} of the form
$R_{jk}= SP(\psi^{\dagger} \sigma_j \psi \sigma_k) =
SP(\psi \sigma_k \psi^{\dagger} \sigma_j)$.

Finally one needs to address a generalized {\it principle of geometric
covariance}, which considers left, right and both-sided operations on
the wavefunction.  This will give interactions which couple to both the
right and left side of the multivector wavefunction.  The generalized
{\it bilateral multivector covariant derivative}
would have the form,

$$ \nabla_\mu (\psi) = \partial_\mu \psi +
 \Omega_\mu^{\ \alpha \beta} \Gamma_\alpha \psi \Gamma_\beta , \eqno(5a)$$
$$ [\nabla_\mu,\nabla_\nu](\psi) =
K_{\mu \nu}^{\ \alpha \beta} \Gamma_\alpha \psi \Gamma_\beta . \eqno(5b)$$

This poses several difficulties in formulation which have yet to be fully
addressed.  For example, eq. (1d) will not give the correct field strength
tensor of eq. (5b) because of the necessity of operating on both sides of
the wavefunction.  However, it is apparent that far more number of gauge
fields can be represented in this way for a given dimension Clifford
algebra than for the spin-gauge type theory.  For example, in the
particular case of the multivector Dirac equation in Majorana algebra,
it was earlier argued that spin gauge theory would yield 10 fields (from
the bar-negative vectors and bivectors of the Clifford group).  The general
bilateral connection of eq. (5a) implies 16x16=256 possible fields.  It has
been
shown \cite{Pezz93} that 136 of these can be interpreted as describing all the
known spin 0 and 1 light unflavored mesons.  Geometric constraints between spin
and isospin suppresses the remaining 120 couplings, which exactly corresponds
to
those mesons forbidden by the standard model (which must appeal to the
requirement of antisymmetry of a composite quark wavefunction).

\section{Summary}
The last category is the most general, containing all the previous categories
as subsets.  It is hence a prototype for the full exploitation of Clifford
algebra in a physical formulation.  Further it allows for a broadening of
the postulates of
physics (n.b. quantum mechanics) beyond mere reformulation.

\acknowledgements
Thanks to Professor Craig Harrison and Victoria Berdon of the Philosophy
Department, and graduate student M. Enjalran of the Physics Department
at San Francisco State University for helpful discussions.  Also to
Professor J.S.R. Chisholm (Univ. of Kent, Canterbury) and J. Crawford
(Pennsylvania State University) for educating the author on the
subject of spin-gauge theories while providing hospitality at their
respective institutions for the author during the summer of 1991.
Finally we thank the University of Ghent, Belgium for the financial
support which made it possible to attend the conference in Deinze.


\begin{thebibliography}{99}

\bibitem {Chis89} Chisholm, J.R.S. and Farwell, R.:1989
{\it J. Phys.} {\bf A22}, 1059.

\bibitem{Craw93} Crawford, J.P.:1993, 'Local Automorphism Invariance:
A Generalization of General Relativity',
{to appear in these proceedings.}

\bibitem{Enja93} Enjalran, M.:1993,'Spin Gauge theory of Pauli Particles',
{Master's thesis, San Francisco State University.}

\bibitem{Grei84} Greider, K.:1984, {\it Found. Phys.}{\bf 14},467;
:1980,{\it Phys. Rev. Lett.}{\bf 44}, 1718.

\bibitem{Pezz592} Pezzaglia, W.:1992, {\it Found. Phys. Lett.}{\bf 5}, 57.

\bibitem{Pezz692} Pezzaglia, W.:1992, 'Generalized Fierz Identities and the
Superselection Rule for Geometric Multispinors',
{Proceedings of the Second Max Born Symposium,
Univ. of Wroclaw, Poland, September 24-27}.

\bibitem{Pezz93} Pezzaglia, W.:1993, 'Dextral and Bilateral Multivector Gauge
Field
Description of Light-Unflavored Mesonic Interactions',{Preprint SFSU-TH-92-03}.

\end{thebibliography}
\end{document}